\documentstyle[aaspp4]{article}

\begin{document}

\title{Orphan GRB radio afterglows: Candidates and constraints on beaming}

\author{Amir Levinson\altaffilmark{1}, Eran Ofek\altaffilmark{1},
Eli Waxman\altaffilmark{2} \& Avishay Gal-Yam\altaffilmark{1,3}}
\altaffiltext{1}{School of Physics and Astronomy, Tel Aviv University,
Tel Aviv 69978, Israel\hfil}
\altaffiltext{2}{Department of Condensed Matter Physics, Weizmann Institute,
Rehovot 76100, Israel}
\altaffiltext{3}{Colton Fellow}

\begin{abstract}

The number of orphan radio afterglows associated with $\gamma$-ray bursts
(GRBs) that should be detected by a flux limited radio survey, is calculated.
It is shown that for jetted GRBs this number is smaller for smaller jet opening
angle $\theta$, contrary to naive expectation.  For a beaming factor
$f_b^{-1}\equiv(\theta^2/2)^{-1}\simeq500$, roughly the value inferred by
Frail et al. (2001) from analysis of afterglow light curves, we predict that between
several hundreds to several thousands orphan radio afterglows should be
detectable (over all sky) above 1 mJy at GHz frequencies at any given time. This
orphan population is dominated by sources lying at distances of a few hundred
Mpc, and having an age of $\sim1$~yr.

A search for point-like radio transients with flux densities greater than 6 mJy was
conducted using the FIRST and NVSS surveys, yielding a list of 25 orphan candidates.
We argue that most of the candidates are unlikely to be radio supernovae.  However,
the possibility that they are radio loud AGNs cannot be ruled out without further
observations.  Our analysis sets an upper limit for the
all sky number of radio orphans, which corresponds to a lower limit $f_b^{-1}>10$ on the
beaming factor. Rejection of all candidates found in our search would imply
$f_b^{-1}>100$. This, and the fact that some candidates may indeed be radio afterglows,
strongly motivate further observations of these transients.

\end{abstract}

\section{Introduction}

Our understanding of gamma-ray bursts (GRBs) has been revolutionized
by the discovery of x-ray afterglows by BeppoSAX (e.g. \cite{Costa97}),
and the following detections of optical transients
(e.g. \cite{Paradijs}).  These efforts led eventually to a confirmation
of the cosmological nature of GRBs, through both direct redshift measurements
(e.g. \cite{Metzger97}), and imaging of the host galaxies (e.g. Sahu et~al. 1997).
The enormous power released in the explosion implies extremely large compactness
of the source and, therefore, most GRB models involve compact or collapsed
objects (e.g. \cite{Goodman86,Bohdan86,Eich89,Woosley93,Lev93,Hypernova98};
see M\'esz\'aros 1999 for a review). In spite of the
impressive successes of expanding relativistic ``fireball'' models
(e.g.\cite{Rhoads93,Katz94,MnR97,Vietri97a,Waxman97a,SPN98}; see
 \cite{MesRev02} for review),
the precise nature of GRB progenitors remains unknown.

The total energy emitted in the GRB explosion and the rate at which such
explosions occur in the Universe are important keys to the understanding
of the progenitors. The determination of these factors is complicated,
however, by virtue of relativistic beaming.  During the phase of $\gamma$-ray
emission the fireball expands with a large Lorentz factor,
$\Gamma\sim10^{2.5}$, so that a distant observer receives radiation from
a conical section of the fireball of opening angle $\sim10^{-2.5}$ around the
line of sight. Thus, estimates of total energy and rate based on
$\gamma$-ray observations are highly uncertain.  The first evidence that the
fireball may be jetted was provided by radio
observations of GRB970508 (\cite{WKF98}).  These observations imply a jet
of relatively wide opening angle, which expands sideways and approaches
%Eli 6/3
sub-relativistic, spherical expansion $\approx1$~yr following the GRB (\cite{FWK00}).
The analysis of radio observations during the sub-relativistic phase allowed
to determine the total GRB energy to be $\approx10^{51}$~erg.
%-----

Jetted GRBs have been widely invoked to explain the optical light curves of the
afterglow emission, most notably so for the source GRB990123
(\cite{Stanek99,Harrison99}). Frail et al. (2001) have analyzed a sample of
GRB afterglows with known redshifts. They find that most bursts are
jetted,
%Eli 6/3
with jet opening angle $\theta\sim0.1$ and average ``beaming factor,''
defined as $f_b^{-1}\equiv(\theta^2/2)^{-1}$, of $f_b^{-1}\simeq500$.
They also find that the total
%-----
$\gamma$-ray energy release, when corrected for beaming as inferred
from the afterglow light curves, is narrowly clustered around
$0.5\times10^{51}$~erg. This result, although somewhat model dependent,
also suggests that the conversion efficiency of fireball kinetic energy to
radiation is high,
in agreement with the conclusion of Freedman \& Waxman (2001), who derived
total fireball energy using early X-ray afterglow data. The energy estimates
of Frail et al. are in agreement with those derived by Freedman \& Waxman, and
also with the total energy derived for GRB970508 (\cite{FWK00}).

%Eli 6/3
Using radio surveys to constrain GRB beaming, rate and energetics
through the search for radio emission from GRBs that were not
necessarily detected in $\gamma$-rays, has been proposed by Perna
\& Loeb (1998) ,
%Eli 14/3
Woods \& Loeb (1999)
%-----
and by Paczy\'nski (2001). Perna \& Loeb have suggested to search
for such "orphan radio afterglow" emission from GRB jets pointing
toward us,
%Eli 14/3
while Woods \& Loeb have suggested to look for $\sim10^3$~yr old,
non-relativistic remnants.
%-----
Paczy\'nski suggested, based on the radio observations of
GRB970508, to search for emission from $\sim1$~yr old nearby GRB
remnants, which have undergone the transition from a relativistic
jet to spherical sub-relativistic expansion, during which emission
is isotropic. Given the hints for an association of GRBs with
supernovae (\cite{Galama98,Bloom99,Reichart99}), it was suggested
by Paczy\'nski to search for strong radio emitters among nearby
supernovae.
%-----

%Eli 6/3
In this paper we use the model, proposed in (\cite{FWK00,LnW00})
for the long term radio afterglow emission of GRBs, to calculate
the expected number of orphan afterglows in radio surveys. We find
that the detectable orphan population is dominated by $\sim1$~yr
old GRB remnants which have just undergone the transition to
sub-relativistic, spherical expansion. We determine the dependence
of the expected number of sources on uncertain model parameters,
derive a lower limit to the expected number of sources, and
demonstrate that independent constraints on the beaming factor of
early GRB emission can be imposed using radio surveys.
%Eli 14/3
We searched for point-like radio transients, by comparing the FIRST
and NVSS wide field catalogues, taken a few years apart, looking
for sources present in one catalogue and absent in the other. Our
approach is different than that proposed by Paczy\'nski, namely we
look directly for radio transients without relying on an
association with nearby supernovae, since the link between GRBs
and luminous supernovae is not well established: The only clear
case, 1998bw, had an atypical (very weak) GRB, while there are
three low redshift, $z\le0.45$, GRBs\footnote{see
http://www.aip.de/~jcg/grbgen.html} with no prominent supernova
reported.
%-----

In \S~2 we provide a brief description of the afterglow model and
its main results used in later sections. In \S~3 we calculate the
number of radio afterglows expected to be detected above a certain
flux. In \S~4 we present a search for point-like radio transients
using the FIRST and NVSS catalogs, and discuss possible
contamination by radio supernovae and AGN.  The implications of
our results are discussed in \S~5.

\section{An outline of the afterglow model}

The basic model assumes an ultra-relativistic, jet like ejecta propagating into an
ambient medium of density $n=1 n_0{\rm\ cm^{-3}}$, slowing
down and expanding sideways, ultimately becoming non-relativistic.
As long as the jet Lorentz factor $\Gamma$ is larger than the inverse of
the jet's opening angle $\theta$, it behaves as if it were a conical section
of a spherical fireball. Once $\Gamma$ drops below
$\theta^{-1}$, the jet expands sideways, and its behavior deviates from that
exhibited in the conical phase (\cite{Rhoads97,Rhoads99}).
After a transition stage, in which the jet expands sideways, the flow
approaches spherical symmetry and becomes sub-relativistic
(\cite{WKF98,FWK00}). The transition from a jet to a
spherical sub-relativistic evolution takes place over a time
scale,
\begin{equation}
t_{SNT}\simeq 6\times10^6(E_{51}/n_0)^{1/3}\ \ {\rm s}
\label{tSNT}
\end{equation}
(\cite{FWK00,LnW00}).  Here, $E_{51}$ is the total energy carried by the jet,
in units of $10^{51}$~erg,
%Eli 6/3
corrected for beaming, i.e. it is related to the isotropic equivalent energy
$E_{\rm iso}$ by
$E=(\theta^2/2)E_{\rm iso}=f_bE_{\rm iso}$,
where $\theta$ is the jet opening angle.
%-----
The subscript SNT refers to the Sedov-von
Neumann-Taylor self-similar solution that describes the sub-relativistic flow.

After the transition, the power is dominated by synchrotron emission
from a sub-relativistic fireball (e.g.\ Frail, Waxman \& Kulkarni 2000).
Denoting by $\xi_e$ ($\xi_B$) the fraction of thermal energy behind the
shock that is carried by electrons (magnetic field), and assuming that
the electrons are accelerated to a power-law energy distribution,
$dn_e/d\gamma_e\propto\gamma_e^{-p}$ with $p=2$, it can be shown (\cite{LnW00})
that the flux at frequencies above the synchrotron peak frequency
at $t=t_{\rm SNT}$,
\begin{equation}
\nu_*\approx1\left({1+z\over2}\right)^{-1}\left({\xi_e\over0.3}\right)^2
\left({\xi_B\over0.3}\right)^{1/2}n_0^{1/2}\,{\rm GHz},
\label{eq:nu_m}
\end{equation}
is given by (see also Appendix of Frail, Waxman \& Kulkarni 2000)
\begin{equation}
f_\nu\approx2h^2\left({1+z\over2}\right)^{1/2}\left({\xi_e\over0.3}\right)
\left({\xi_B\over0.3}\right)^{3/4}\left(\frac{d_{L}}{R_0}\right)^{-2}n_0^{3/4}E_{51}
\nu_{9}^{-1/2}\left[{t\over t_{\rm SNT}(1+z)}\right]^{-9/10}\,{\rm
mJy}~~.
\label{eq:f_nu}
\end{equation}
Here, $d_L$ is the luminosity distance, $R_0=c/H_0=10^{28}h^{-1}$ cm, and $\nu_9$ is the
observed frequency measured in GHz.
%Eli 6/3
Analysis of GRB970508 radio data implies
(\cite{FWK00}) $n_0\approx1$, $\xi_e\sim\xi_B\sim1/3$.
%-----

\section{Expected number of orphan afterglow sources in radio surveys}
\label{sec:Expected}

Consider a survey with a flux threshold denoted by $f_{\nu min}$, and
denote by $\tau_m$ the time, in units of $t_{SNT}(1+z)$, at which
the radio flux of a GRB at a given luminosity distance, $d_L$, drops below
detection limit.  From eq. (\ref{eq:f_nu}) we then obtain
\begin{equation}
\tau_m=\eta (1+z)^{5/9}(d_L/R_0)^{-20/9},
\label{tau}
\end{equation}
where
\begin{equation}
\eta=h^{20/9}\left(\frac{f_{\nu min}}{1 mJy}
\right)^{-10/9}\left(\frac{\xi_e}{0.3}
\right)^{10/9}\left(\frac{\xi_B}{0.3}\right)^{10/12}
\left(\frac{2}{\nu_9}\right)^{5/9}n_0^{10/12}E_{51}^{10/9}.
\label{eta}
\end{equation}

%Eli 6/3
We shall assume, in what
follows, that the total energy output and beaming factor of GRBs are standard
and, moreover, that the fraction of total energy release that goes into driving
the external blast wave generating the afterglow emission, is the same for all
GRBs. We make this assumption in order to simplify our analysis, and since
the distribution of the above parameters is not well known.
It is important to emphasize at this point that,
as mentioned above, the analysis presented in Frail et al. (2001)
indicates that the distribution
of GRB true (i.e. corrected for beaming) gamma-ray energies
is narrowly clustered around $E_{\gamma,51}\simeq 0.5$,
with a corresponding average beaming
factor of $<f_b^{-1}>=<({\theta^2}/2)^{-1}>\simeq 500$.
These results also imply high, order unity, radiative efficiency,
consistent with the result obtained
by Freedman \& Waxman (2001) using a different method, and implying
that the true total fireball energy is a few times $E_{\gamma}$, i.e.
narrowly peaked around $E_{51}\simeq 1$. Thus, we expect corrections
to our results, due to deviations of parameters from their standard values,
to be small. Moreover,
our results, given as explicit functions of the standard
parameter values, can be easily generalized to the case where parameter values
are drawn from a general distribution, by replacing these values
with their corresponding averages.
%-----

Let $\tau_i$ be the time at which the radio source just becomes isotropic.
Then for a given flux threshold, the maximum luminosity distance below
which a substantial number of sources can be detected
is obtained from eq. (\ref{tau}) for $\tau_m=\tau_i$:
\begin{equation}
d_{Lmax}(z_{max})=R_0(\eta/\tau_i)^{9/20}(1+z_{max})^{1/4}.
\label{dL}
\end{equation}
For a given choice of GRB parameters, a flux threshold $f_{min}$ at the observed
frequency $\nu_9$, and value of $h$, the parameters $\eta$ and $\tau_i$ are
fixed.  Equation (\ref{dL}) can then be numerically
solved for the assumed cosmology to yield the maximum redshift $z_{max}$
encompassed by the survey.

We denote by $\dot{n}$ the observed number of GRB events per year
per Gpc$^3$ at $z=0$ (which, assuming that the GRB rate follows the redshift
evolution of the star formation, is estimated to be $\dot{n}\approx0.5$ (\cite{Schmidt01})),
and by $\Phi(z)$ the redshift evolution factor.
The expected number of radio afterglows between $z$ and $z+dz$ can then
be expressed as,
\begin{equation}
\frac{dN_R}{dz}=N_0d_L^2 \frac{dl}{dz}(\tau_m-\tau_i)\Phi(z),
\label{dN}
\end{equation}
where $dl=R_0^{-1}cdt$ is the light distance in units of $R_0$, and
\begin{equation}
N_0=f_b^{-1}4\pi R_0^3\dot{n} t_{SNT}
%Eli 6/3
\simeq 3.5\times10^4 (500f_b)^{-1}h^{-3}\dot{n}
%-----
(E_{51}/n_0)^{1/3}.
\label{N0}
\end{equation}
The total number expected, $N_R$, can be obtained by integrating eq
(\ref{dN}) from
$z=0$ to $z=z_{max}$, with $z_{max}(\eta/\tau_i)$ determined from eq. (\ref{dL}).

As an example, consider a survey with a flux threshold
of 5 mJy.  Assuming $\tau_i=3$, $h=0.75$ and the above choice for the
remaining parameters, we find $d_{Lmax}\simeq 0.2R_0=0.8$ Gpc,
and a corresponding redshit $z_{max}\simeq 0.2$.
At such small redshifts cosmological effects can be neglected to a reasonable approximation,
and the luminosity distance can be taken to be equal to the proper distance.  In this case
eq. (\ref{dN}) can be integrated analytically.  One then finds, using
$n=10^{-1}n_{-1}{\rm cm}^{-3}$
%Eli 6/3
\begin{eqnarray}
N_R&&\simeq N_0\eta^{27/20}\tau_i^{-7/20}
\cr&&\simeq 18(500f_b)^{-1} (\dot{n}/0.5)\left(\frac{f_{\nu min}}{5 mJy}
\right)^{-3/2}\left(\frac{\xi_e}{0.3}
\right)^{3/2}\left(\frac{\xi_B}{0.03}\right)^{9/8}n_{-1}^{19/24}E_{51}^{11/6}
\nu_9^{-3/4}(\tau_i/3)^{-7/20}
\cr&&\simeq
10^{4}f_b^{5/6} (\dot{n}/0.5)\left(\frac{f_{\nu min}}{5 mJy}
\right)^{-3/2}\left(\frac{\xi_e}{0.3}
\right)^{3/2}\left(\frac{\xi_B}{0.03}\right)^{9/8}n_{-1}^
{19/24}E_{\rm iso,54}^{11/6}\nu_9^{-3/4}(\tau_i/3)^{-7/20}.
\label{N}
\end{eqnarray}
As seen, $N_R$ depends only weakly on $\tau_i$.  Consequently, any
uncertainty in the time at which the radio emission becomes
roughly isotropic should not affect our result considerably. We
have also given the result in terms of the isotropic equivalent
energy, $E_{\rm iso}$, for which we have chosen a normalization of
$10^{54}$~erg, since for the sample of GRBs with known redshifts
given in Frail et al. 2001, one finds $<E_{\rm \gamma,
iso}>=3\times10^{53}$~erg and $<E_{\rm \gamma,
iso}^{11/6}>^{6/11}=5\times10^{53}$~erg. Since the true energy
inferred from the observations, which is constrained by the
isotropic equivalent energy $E_{\rm \gamma, iso}$ measured in GRBs
with known redshifts, is inversely proportional to the beaming
factor, $N_R\propto f_b^{5/6}$. This means that the number of
orphans expected to be detected in a flux limited survey decreases
with increasing beaming factor, unless the survey is sufficiently
deep to detect all sources.
%Eli 14/3
This is, of course, due to the fact that larger beaming factor
implies, for given isotropic equivalent energy, a lower true
energy, and hence a lower radio flux at the transition to
spherical expansion, as demonstrated by Eq. (\ref{eq:f_nu}).
%-----

The number of orphan radio GRBs, Eq. (\ref{N}), depends on the model parameters
$E$, $n$, $\xi_e$ and $\xi_B$. Afterglow observations strongly suggest that
$\xi_e$ is close to equipartition, i.e. $\xi_e\sim1/3$ (Freedman \& Waxman 2001),
and that $E$ is narrowly peaked around $E_{51}=1$ (Frail et al. 2001).
The main uncertainty is thus in the parameters $n$ and $\xi_B$.
The peak flux of a GRB afterglow, for
an observer lying along the jet axis, is proportional to
$E_{\rm iso}\sqrt{n\xi_B}$, and for typical luminosity distance of
$3\times10^{28}$~cm it is $\approx10\sqrt{\xi_B n_0/10^{-3}}E_{\rm iso,54}$~mJy
(\cite{Waxman97b,GruzinovWaxman99,WijersGalama99}). Observed afterglow fluxes
generally imply $\xi_B n_0\ge 10^{-3}$ for $E_{\rm iso,54}\sim0.1$,
and values $\xi_B n_0\sim 1$ are obtained in several cases
(e.g. \cite{Waxman97b,WijersGalama99,KumarPan01}). Our parameter choice in
Eq. (\ref{N}), $(\xi_B/0.3)=0.1$ and $n_0=0.1$, is therefore conservative.
Even for such a conservative choice of parameters we expect more than a
dozen radio afterglows to be detected by an all sky survey like the NVSS,
provided the beaming factor
is not much larger than that anticipated from other considerations.
Note, that our choice $\dot{n}=0.5$ is also conservative: If the GRB rate does
not evolve with redshift, rather than following the star formation rate,
the $z=0$ rate is about 8 times larger, $\dot{n}\approx4$, and the expected
number of radio GRB remnants is correspondingly larger by the same factor.
Finally, the brightest source should have a flux of
\begin{equation}
f_{\nu max}\simeq 30(500f_b)^{-2/3}
(\dot{n}/0.5)^{2/3}\left(\frac{\xi_e}{0.3}
\right)\left(\frac{\xi_B}{0.03}\right)^{3/4}n_{-1}^{19/36}E_{51}^{11/9}
\nu_9^{-1/2}(\tau_i/3)^{-7/30}\ \ {\rm mJy}.
\label{fmax}
\end{equation}
Since the flux in this phase declines roughly as $(t/t_{SNT})^{-1}$, such a source
should remain above 1 mJy for at least several years.  This motivates follow-up
observations of the brightest transients found in a survey.

The derivation of eq. (\ref{N}) does not take into account the contribution from sources
having a lifetime shorter than $\tau_i t_{SNT}$.  These sources are expected to be
beamed, and so the fraction that can be observed (those directed along our sight line)
should be of course
correspondingly smaller.  On the other hand, the radio flux emitted at times earlier than
$\tau_i t_{SNT}$ is larger and, consequently, these beamed sources can be seen out to a larger
distance.  We may estimate the number of beamed radio sources anticipated to be
detected by the survey using the following argument.  The ratio of
the number of sources having an opening angle $\sim\theta$ that are pointing in our
direction, to the number of sources that had just become isotropic, is given by
$(2\pi\theta^2/4\pi)[T(\theta)/t_{SNT}]$ for a two sided jet, where $T(\theta)$
is the observed expansion time to opening angle $\theta$:
$T(\theta)\propto\theta^2$ (e.g. \cite{Rhoads99}).
Noting that the radio flux measured by an observer
within the beam drops as $t^{-1/3}$ during the expansion of the jet (e.g. \cite{Rhoads99}),
the flux of a beamed source is higher than that of a source that just became spherical (lying
at the same distance) by a factor $\theta^{-2/3}$. This implies that beamed sources can be
seen out to a distance larger by a factor $\theta^{-1/3}$ compared to those that just became
spherical. Thus, the ratio of the number of beamed sources to those that just became
spherical, observed above a certain flux, is
$(\theta^{-1/3})^3(\pi\theta^2/4\pi)[T(\theta)/t_{SNT}]\propto\theta^3$. Thus, the number
of detectable sources is dominated by sources which have already expanded to spherical
symmetry.

For deeper surveys cosmological effects cannot be ignored and must be taken into
account.  We have solved eqs. (\ref{dL}) and (\ref{dN}) numerically for different
choices of cosmological parameters.  For the range of parameters presently
favorable we find that $N_R$ depends only weakly on the assumed cosmology.
In the following we present the results obtained for $h=0.75$, $\Omega_M=0.3$,
and $\Omega_{\Lambda}=0.7$.  We also made the popular assumption that the
comoving GRB density distribution traces the star formation rate, and
invoked a strong redshift evolution for the later, as suggested by
Madau \& Pozetti (2000).  Specifically, we take $\Phi(z)=(1+z)^3$ in
the following calculations.

The solution of eq. (\ref{dL}) is depicted in fig 1a, where the dependence of the
maximum redshift $z_{max}$ on $\eta/\tau_i$ is plotted.
Fig. 1b shows the dependence of $N_R/N_0$ on $\eta$ for different values of $\tau_i$.

\begin{figure}[h]
%\plottwo{z.eps}{NN2.eps}
\centerline{\epsfxsize=120mm\epsfbox{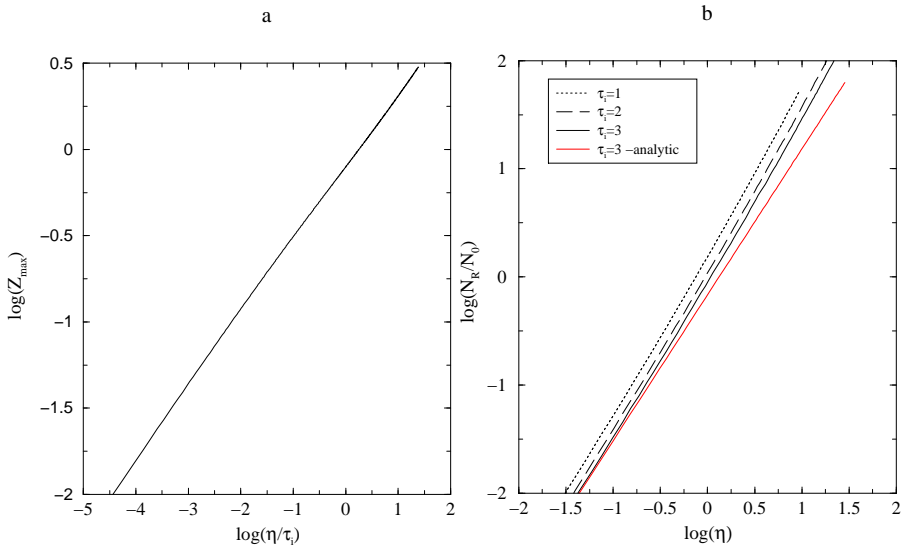}}
\caption{(a) A plot of the maximum redshift below which sources
can be detected above a certain flux, $f_{\nu min}$, versus
$\eta/\tau_i$. The parameters $\eta$ and $\tau_i$ are defined in
the text. For fixed afterglow parameters $\eta\propto f_{\nu
min}^{-10/9}$. (b) The cumulative number of radio afterglows
$N_{R}$ above a certain flux, normalized to $N_0$,  versus $\eta$,
for different values of $\tau_i$.  the normalization factor $N_0$
is given by eq. (\ref{N0}) in the text.  The analytic solution
obtained in the Euclidean case in eq. (\ref{N}) for $\tau_i=3$ is
displayed (red line) for a comparison } \label{z}
\end{figure}

For the cosmology adopted above, the radio flux of a source at a redshift of
$z=2$ is given by
\begin{equation}
f_{\nu}\simeq 50 \left(\frac{\xi_e}{0.3}
\right)\left(\frac{\xi_B}{0.3}\right)^{3/4}n_0^{3/4}E_{51}\nu_9^{-1/2}
\left(\frac{t}{t_{SNT}}\right)^{-9/10}\ \ {\rm \mu Jy}
\end{equation}
and is smaller by a factor of 4 at $z=3$.  Thus for our above (conservative) choice of
afterglow parameters a flux limit of $\le 1 \mu Jy$ is required
to see all the orphans if the GRB population is dominated by those at $z\le 2$.

\section{A Search for NVSS/FIRST Radio Transients}

\subsection{Analysis}
\label{sec:Analysis}

We conducted a search for point-like radio transients
in the FIRST radio catalog (White et al. 1997).
We searched for compact FIRST radio sources
that do not appear in the NVSS
radio catalog (Condon et al. 1998).
The sample we obtain this way is incomplete, since
it rejects afterglow sources that were detected
by the NVSS, and were bright enough to remain above 6 mJy
during the FIRST observation.  We examine this incompleteness factor
further below and show that it is smaller than 5 percent.
As the FIRST and NVSS surveys were obtained
in different VLA configurations,
the comparison was done carefully.
We use this comparison to put an upper limit on the
number of GRB radio afterglows.

The FIRST is a 20cm radio survey conducted using
the NRAO VLA in B-configuration,
with $5''$ beam-size.
%The survey utilizes $2\times7$ $3$-MHz frequency
%channels centered at $1365$ and $1435$~MHz.
The final maps have a typical noise RMS of $0.15$~mJy.
This value allows $5\sigma$ detection of $1$~mJy sources.
The survey's resolution enables an astrometric accuracy
better than $1"$ ($90\%$ CL).
The FIRST radio survey has begun in 1994,
and when completed it will cover $10,000$~deg$^{2}$,
mostly in the north Galactic cap.
In the search described below, we used
the FIRST 2001, Oct 15 version,
covering $8,565$~deg$^{2}$.
This catalog has $771,076$ sources,
hence, a density of about $90$ sources per deg$^{2}$.

The NVSS 20cm radio survey (Condon et al. 1998) was
conducted between 1993 to 1996
and covers all the sky north of declination $-40$~deg.
It utilized the VLA D- and DnC-configurations,
with a circular beam-size of $45''$.
This is significantly larger than the median angular size ($\sim10''$)
of faint extragalactic sources.
With this beam-size, the faintest NVSS sources
have an astrometric accuracy better than $7"$ ($68\%$ CL).
The images have RMS brightness fluctuations of $\sim0.45$~mJy~beam$^{-1}$,
that allows for $5\sigma$ detection of $\sim2.5$~mJy sources.
The NVSS $99\%$ completeness level is about $3.4$~mJy.
With density of $\rho\sim60$ sources per deg$^{-2}$,
it has total of $\sim2\times10^{6}$ radio sources.

\begin{figure}[h]
\centerline{\epsfxsize=120mm\epsfbox{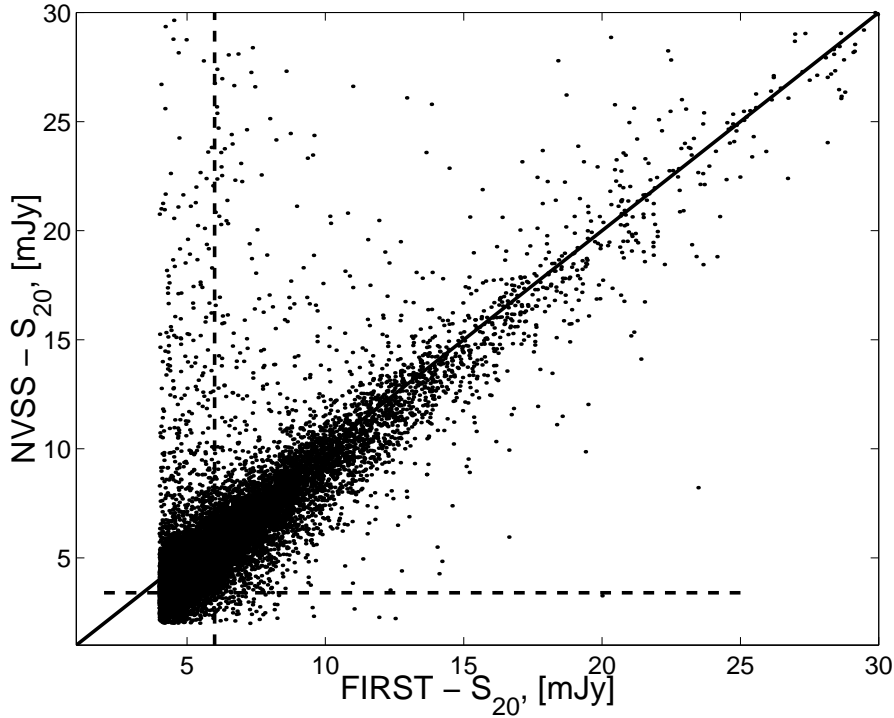}}
\caption {FIRST vs. NVSS peak flux for all FIRST
sources with NVSS counterparts.
The solid line shows the location where the FIRST/NVSS flux ratio equals 1,
the horizontal dashed line shows the NVSS $99\%$ completeness
flux level of $3.4$~mJy, and the vertical dashed line shows
the $6$~mJy cut.}
\label{FIRST_NVSS_FLUX}
\end{figure}

Radio sources could be intrinsically variable.
Therefore, using the NVSS $99\%$ completeness of $3.4$~mJy
as our threshold, could introduce a large number
of variable sources
(e.g., that were detected in the FIRST survey, but were
just under detection in the NVSS)
into our transients list.
In order to compare the two catalogs, we need to choose
a flux limit that rejects
most of the variable objects.
We plot in
Figure~\ref{FIRST_NVSS_FLUX} all FIRST point-like
sources (above $4$~mJy) with NVSS counterparts.
The solid line shows the location of sources for which the
FIRST flux equals the NVSS flux,
and the horizontal dashed line shows the NVSS, $99\%$ completeness
flux level (i.e., $3.4$~mJy).
Based on this plot we chose a threshold of $6$~mJy,
for which the scatter around the solid line
does not produce large number of false transients sources.

In the following we assumed that all the FIRST images,
were taken after the corresponding NVSS images.
This assumption
will be examined more carefully later.
Unless a
GRB afterglow happens in a radio galaxy,
it is expected to:
(i) be a point source;
and (ii) be absent from the NVSS.
Since radio galaxies are a tiny fraction of the overall galaxy population,
these assumptions hold for most of the afterglow population.

We selected all the FIRST point-like sources with peak flux
above $6$~mJy
($21.5\%$).  From this subsample, we selected all the
unresolved objects ($4.3\%$ of the $6$~mJy subsample).
However, some of the resolved FIRST objects
could actually be point sources, classified as resolved
due to measurements errors.
Adopting the FIRST size error criterion (White et al. 1997),
for $9.2\%$ of the FIRST sources above $6$~mJy
it is not possible to reject the hypothesis that they
are point-like at the $99.7\%$ CL.
Therefore, the number of sources above $6$~mJy
for which the point-source hypothesis
could not be rejected at the $3\sigma$ level
is $2.1$ ($=9.2/4.3$) times larger than the number of
point sources we used in our sample.
We ended with a catalog of $7181$ FIRST point-like sources
with peak flux above $6$~mJy.
For these sources, we searched the NVSS for counterpart
within $30"$ of the FIRST position.
Although the NVSS astrometry is better than $7"$,
the use of such a high threshold was necessary because
in cases of two (or more) nearby point-like radio sources,
the NVSS could detect only one elongated source,
with its center offset by more than $7"$ from one
of the FIRST point source positions.
The search yields
$110$ point-like FIRST sources with peak flux above $6$~mJy
for which there is no NVSS counterpart.

We examined the POSS-II $E$ images
and the FIRST and NVSS radio maps of all these transient
candidates.
We found that: $2\%$ result from holes in the NVSS maps;
$14\%$ are artifacts due to
un-cleaned aliases in the radio maps;
$61\%$ are multiple sources for which
the NVSS found an elongated source centered more than $30"$
from one of the FIRST radio positions;
and for the rest ($23\%$, $25$ sources) we could not
reject the hypothesis that these are radio transients
or variable radio sources
of some kind (e.g., AGNs, SNe, GRBs, Pulsars).
A large fraction of these 'sources' are possibly:
radio artifacts; NVSS non-detections due to locally high background
(e.g., in the vicinity of bright radio source);
or physical phenomena other than GRBs (e.g., AGNs, SNe).
If we change the flux limit from $6$ to $10$~mJy,
the number of candidates decreases from $21$ to $6$.

The following table lists the $25$ candidates along with their basic
properties.

\begin{table}[t]
\label{CandTable}
\caption{A list of radio-transient candidates.}
\vspace{0.2cm}
\begin{centering}
\begin{minipage}{140mm}
\begin{tabular}{ccccl}
\tableline
R.A. & Dec. & FIRST radio & Quality$^{a}$ & Optical counterparts and comments \\
(J2000) &   & flux [mJy]  &               &                                   \\
\tableline
12:56:32.34 & -05:47:39.7 &  9.8 & 2 & No optical source                      \\
13:07:13.39 & -05:27:09.4 &  6.6 & 2 & No optical source                      \\
12:26:07.68 & +02:02:52.4 & 12.9 & 1 & No optical source                      \\
12:28:47.38 & +02:03:12.8 & 34.5 & 3 & No optical source                      \\
12:29:02.49 & +02:31:03.1 &  9.6 & 2 & No optical source                      \\
12:26:50.43 & +02:39:32.6 &  6.8 & 1 & Blue point source with $R\sim19$ mag   \\
10:08:02.94 & +07:33:18.2 &  6.5 & 3 & No optical source                      \\
12:30:23.41 & +11:31:39.0 &  7.6 & 2 & Point source with $R\sim18$ mag        \\
12:25:32.62 & +12:25:00.5 &  7.2 & 2 & Nucleus of an $R\sim16$ galaxy         \\
12:29:00.02 & +12:48:18.8 &  6.2 & 2 & No optical source                      \\
12:15:50.24 & +13:06:54.0 &  9.7 & 2 & Located in an NGC4216 galactic arm     \\
08:21:50.18 & +17:46:16.3 &  6.0 & 2 & No optical source                      \\
11:43:55.34 & +22:10:20.4 &  6.2 & 2 & No optical source                      \\
16:52:03.08 & +26:51:39.9 &  6.3 & 2 & Coincident with the PSR J1652+2651     \\
13:31:14.55 & +30:25:57.8 & 10.6 & 3  & No optical source $^{b}$              \\
13:31:18.21 & +30:26:01.1 & 10.8 & 3  & No optical source $^{b}$              \\
13:31:01.86 & +30:33:00.8 & 13.1 & 3  & No optical source $^{b}$              \\
13:31:14.42 & +30:55:39.2 &  8.7 & 2  & No optical source $^{b}$              \\
13:31:17.11 & +31:01:53.2 &  7.4 & 2  & No optical source $^{b}$              \\
13:24:20.39 & +31:41:37.2 &  9.2 & 1 & $0.2'$ from an $R\sim16$ galaxy        \\
13:25:17.11 & +31:51:07.6 &  9.4 & 2 & No optical source                      \\
17:20:59.90 & +38:52:26.6 &  8.2 & 2 & No optical source                      \\
15:22:48.69 & +54:26:44.1 &  6.6 & 2 & Point source with $R\sim19$            \\
10:48:48.92 & +55:15:08.7 &  6.1 & 2 & Blue point source with $B\sim21$       \\
08:36:00.07 & +55:55:17.9 & 10.7 & 2 & No optical source                      \\
\tableline
\end{tabular}
Notes:\\
$^{a}$ radio images quality: 1 - probable radio transient,
2 - suspect detection,\\
3 - probably an artifact\\
$^{b}$ group of 'sources' near bright radio source 3C286 - probably
artifacts.
\end{minipage}
\end{centering}
\end{table}

For each of these objects, we conducted a search for known counterparts within
a radius of $5'$ in the NED
database and within $1'$ in the HEASARC database.
Our findings are listed in the table.

$2575$~deg$^{2}$ of the FIRST survey were taken
before the completion of the NVSS survey.
Therefore, we reject all the sources
in this area (i.e., $22.2^{\circ}<\delta<42.5^{\circ}$).
Omitting this strip (PSR J1652+2651 is in this strip), the area covered by our search
is therefore $5990$~deg$^{2}$,
and the number of sources is: $16$ and $3$,
for the $6$~mJy and $10$~mJy flux thresholds respectively.
The upper $95\%$ Poisson CL on these numbers
are: $26$ and $9$ respectively.
We consider these numbers to be conservative
upper-limits on the number of possible GRB afterglows.

Finally, we have to correct our upper limits for the various completeness
factors involved:
(i) Missing point sources due to FIRST measurements errors.
As we showed above, a very conservative estimate for this number is
$C_{missing}=9.2/4.33 = 2.1$;
(ii) Area completeness (e.g., the fraction of the sky covered by our survey)
$C_{area}=41252/5990 = 6.8868$;
(iii) Given The crude NVSS resolution,
the probability that a transient source will be
detected up to $r=90''$ (twice the NVSS FWHM), from
a known un-related NVSS source,
and therefore could be confused with it, is
$P(<r)=1-\exp{(-\pi\rho r^{2})}$
$=11\%$.
Therefore, the source confusion completeness factor is: $C_{conf}=1.11$;
(iv) NVSS completeness - For about $2\%$ of the FIRST sources
there are no corresponding NVSS images (holes in the catalog),
hence we set $C_{comp}=1.02$.
(v) Incompleteness due to our search criteria - afterglow sources detected by
the NVSS that were bright enough to remain above 6 mJy during the FIRST
observation are rejected by our search.  The number of such sources depends on
the average time separation
between the NVSS and FIRST observations.  Denoting by $t_{sep}$ the
time interval between two consecutive observations of a typical afterglow source, eq.
(\ref{eq:f_nu}) implies that the flux during the first observation was larger than during
the second observation by a factor
of $(t_{sep}/t_{SNT}+1)^{9/10}$, assuming the source was near maximum flux during the
first detection (i.e., its age was roughly $t_{SNT}$).  Consequently, the fraction
of bright afterglow sources in the NVSS catalog that may be detected above 6 mJy
by the FIRST at time $t_{sep}$ after 1996, must be smaller than $\delta(t_{sep})
=(t_{sep}/t_{SNT}+1)^{-27/20}$.
The smallest time separation between the catalogs in the strips we consider is about
$t_{min}=6$ months, and the largest is about $t_{max}=4.5$ years.  Taking the
FIRST detections to be uniformly distributed in time, integrating the fraction
$\delta(t_{sep})$ from $t_{min}$ to $t_{max}$ yields a total fraction
of less than 5\% that our search rejected.  Thus, we set $C_{search}=1.05$.
Taking together all these factors, we can put a conservative $95\%$ CL upper
limit of $26 \times C_{area} \times C_{missing} \times C_{conf} \times C_{comp}
\times C_{search}= 447 $
for the all-sky number of GRB afterglows with peak specific-flux
above $6$~mJy (or $154$ for the $10$~mJy threshold).

\subsection{Radio Supernovae}
\label{sec:RSNe}

The characteristic behavior of radio supernovae (RSNe) (see \cite{Weiler02} for a recent
review) appears to be similar to the expected behavior of radio afterglows (RGRBs).
In particular, the radio light curve of type II RSNe rises on
a characteristic time scale of $\sim100$~days (\cite{Weiler98}), comparable to
$t_{SNT}$ (see eq. [\ref{tSNT}]), and has an overall similar shape to that of RGRBs
following the peak.
This is not at all surprising since both are produced by the same
mechanism, namely a blast wave, albeit under somewhat different conditions.
This similarity would render any attempt to distinguish RGRBs from RSNe difficult
and, therefore, motivates a careful examination of potential differences between
these two classes of objects.

We identify two features that may allow one to distinguish RGRBs from RSNe.
Firstly, the radio flux is anticipated to decline smoothly with a
$t^{-1}$ dependence during the sub-relativistic phase in RGRBs, while the
radio light curves seen in RSNe are often
much more complex.  However, some RSNe do show a smooth decline, and those may be
confused with RGRBs.  Secondly, the absolute peak radio luminosity of RSNe may differ
considerably from the radio luminosity of the brightest RGRBs.
For a typical GRB the energy injected into the blast wave is
$E\simeq10^{51}$~erg, and with the above choice of
parameters: $n=0.1 {\rm cm}^{-3}$, $\tau_i=3$, $\xi_B=0.03$ and $\xi_e=0.3$,
the RGRB luminosity density at 5~GHz is $7\times10^{28}{\rm erg\ s^{-1} Hz^{-1}}$ at a time
of $\approx0.5$~yr after the explosion.  This luminosity is
$\sim 100$ times larger than the typical peak luminosity (of detected)
RSNe, of order $10^{27}{\rm erg\ s^{-1} Hz^{-1}}$ (\cite{Weiler98}), and
several times brighter than
the brightest known RSNe, SN1988z, which had a peak flux density
of $2\times10^{28}{\rm erg\ s^{-1} Hz^{-1}}$.  This suggests that in a flux limited
survey RSNe and RGRBs should have a distinct redshift distribution, with RSNe
having preferentially smaller redshifts.

We estimate below the number of RSNe expected to be detected in
a search of the type described in this paper for RGRBs, and compare their
expected redshift distribution to that calculated for RGRBs. In order to
make an accurate prediction for RSNe, their (peak) luminosity function and
the shape of the corresponding light curves must be known.  Unfortunately,
due to the small sample of RSNe presently available (12 type II and 4 type I),
and the lack of a supernova model that is capable of predicting the distribution
of RSN properties, the determination of the peak luminosity function is highly uncertain.
We therefore provide below a crude estimate based on the Weiler et al. (1998)
compilation of RSNe radio data. Based on Figure 3 of Weiler et al., we adopt
a RSNe peak luminosity function with equal number of RSNe per logarithmic
peak luminosity interval over the 5~GHz peak luminosity range
$L_1=10^{23}{\rm erg\ s^{-1} Hz^{-1}}$ to
$L_2=10^{28}{\rm erg\ s^{-1} Hz^{-1}}$.  In addition, we invoke a correlation between
5~GHz peak luminosity $L_p$ and peak time $T_p$ of the form,
$L_p=10^{28}(T_p/10^3{\rm day})^{1.4}{\rm erg/s Hz}$, as demonstrated in Weiler et al. (1998).
These assumptions hold, based on the available data, for type II RSNe. The
properties of Type I RSNe are less certain. However, since type II
RSNe dominate the (detected) RSNe population, and since the type I RSNe
peak luminosities are close to the typical type II peak luminosities, we
expect our estimate below of the type II RSNe ``background'' rate to provide
a (crude) estimate for the total RSNe background.

Assuming that the RSNe flux drops
as $T_p/t$ for $t>T_p$, the number of 5 GHz sources on the sky above a certain
flux density $f_{\nu,\rm min}$ at any given time can be readily computed:
\begin{equation}
 N_{RSNe}=\frac{112\pi}{93\ln(L_2/L_1)}R_{RSNe}^3 \dot n_{RSNe}T_2.
\label{eq:N_RSNe}
\end{equation}
%Eli 6/3
Here, $\dot n_{RSNe}$ is the RSNe rate per unit volume, $T_2$ is the peak
time corresponding to peak luminosity $L_2$,
%-----
and $R_{RSNe}$ is the maximum distance at which a RSN can be observed,
\begin{equation}
  R_{RSNe}\equiv\left(\frac{L_2}{4\pi f_{\nu,\rm min}}\right)^{1/2}=
   41\left(\frac{L_2/10^{28}{\rm erg/s Hz}}{f_{\nu,\rm min}/5{\rm mJy}}
   \right)^{1/2}{\rm Mpc}.
\label{eq:R_RSNe}
\end{equation}
Since the typical distances of detectable RSNe are $\sim10$~Mpc,
we have assumed Euclidean geometry for the calculation above.
It is also straight forward to calculate the average distance of detected
RSNe,
\begin{equation}
  <R>_{RSNe}=\frac{93}{304} R_{RSNe}=
   13\left(\frac{L_2/10^{28}{\rm erg/s Hz}}{f_{\nu,\rm min}/5{\rm mJy}}
   \right)^{1/2}{\rm Mpc}.
\label{eq:Rav_RSNe}
\end{equation}
These distances should be compared with those obtained for RGRBs: for the
conservative GRB parameters quoted above, the maximum and average
distances of RGRBs detectable above 5~mJy at 5~GHz are $R_{GRB}=140$~Mpc
and $<R>_{GRB}=R_{GRB}/3=46$~Mpc respectively.

Using the above equations, and taking $\dot{n}_{RSNe}=1.2\times10^{5} {\rm\ Gpc^{-3} yr^{-1}}$,
corresponding to the total rate of
core-collapse SNe (\cite{Filippenko01}), we find
$N_{RSNe}=3(f_{\nu,\rm min}/5{\rm mJy})^{-3/2}$. Comparing this result with
the estimates for RGRBs given in \S2, we find that the number of
RGRBs detected in a survey of the type described here
should be at least comparable, and most likely significantly
larger, than the number of detected RSNe.
The characteristic distances of detectable RGRBs should be $\ge4$ times larger
than those expected for RSNe.

The anticipated proximity of RSNe in the FIRST survey implies that the host galaxy
should be easily detectable.  In fact, even very faint galaxies should be detectable
above the POSS-II limit out to a distance of 100 Mpc.
This means that all the radio transients listed in table 1 for which no optical
counterpart is detected are unlikely to be RSNe, as even dwarf hosts should be
clearly evident.

\subsection{Radio loud AGNs}
\label{sec:AGN}

Radio loud AGNs are known to exhibit large amplitude variations over a broad range
of timescale and frequencies, with flare durations ranging from minutes to years.
The characteristic durations of radio outbursts (with the exception of ID variability)
are between weeks to years (\cite{Val99,venturi01}), with
amplitudes that occasionally exceed a factor of 10.  Thus, it is conceivable that
some of the transients in our selected sample are radio loud AGNs.  Indeed, our
initial list of 110 transients contains a number of such sources.

Most of the candidates in table 1 have no optical counterpart brighter than $M_R=21$.
What is the likeliness that some or most of them are in fact AGNs?
One way to characterize the radio loudness is in terms of the radio-to-optical luminosity
ratio, $R$, which is commonly defined using radio and optical luminosities at some fiducial
rest-frame wavelengths.  Existing samples of AGNs exhibit a distribution
in radio loudness that depends on selection criteria.  The distribution of $R$ in the
FBQS (\cite{FBQRS}) for example peaks at $R\sim10$, with very few
sources (0.3 \%) having $R>10^{4}$.  The distribution of radio selected quasars
has a median at $R\sim 10^3$ (\cite{Wad99}).  It is not clear at present
whether this is a result of a selection effect, and whether there exists a population of
extremely radio loud AGNs.  However, based on our current knowledge, we would
naively expect that deep enough observations at
optical wavelengths would be able to ultimately reveal the underline AGN.  future X-ray
detections of some of our candidates will also imply that these are likely AGNs.  A precise
determination of the minimum value of $R$ for the objects in table 1 requires a knowledge
of their redshifts, which are not available.  If $R$ is defined as in
Wadadekar \& Kembhavi (1999) then a rough
estimate yields $R>300$ for those objects.  In addition, they should have no extended
structure above $\sim 1 mJy$ on scales larger than a few arcseconds.

Another potential difficulty for future surveys may arise due to the presence of faint,
BL Lac type sources.  If a population of radio BL Lacs exist for which
the continuum flux associated with the nuclear activity drops below that of the host galaxy
during quiescent periods, then a radio transient associated with such an object may be
confused with a RSN or orphan afterglow if optical observations were made when the AGN was in
a low state.  An example is the source SDSS J124602.54 which has been proposed as a candidate
optical orphan afterglow, and later confirmed to be a BL Lac (\cite{Gal02}).  In order
to select out such sources in future surveys, monitoring in the radio and/or
optical bands may be necessary.

\section{Discussion}

We have calculated the number of orphan radio afterglows expected to be detected in
a flux limited survey. For the GRB parameters derived by Frail et al. (2001);
total jet energy $E\simeq10^{51}$~erg and average beaming factor
$<f_b^{-1}>\equiv(\theta^2/2)^{-1}\simeq 500$, the number of sources expected to be
present over all sky at any given time with flux exceeding $f_{\nu\rm min}$ is
[see Eq. (\ref{N})]
\begin{equation}
N_R\simeq 20\left(\frac{\dot{n}}{0.5{\rm Gpc}^{-3}{\rm yr}^{-1}}\right)
\left(\frac{f_{\nu min}}{5\rm mJy}\right)^{-3/2}
\left(\frac{\xi_B}{0.03}\right)^{9/8}n_{-1}^{19/24}.
\label{eq:N_R}
\end{equation}
Here $\dot{n}$ is the GRB rate at redshift $z=0$, $n=0.1n_{-1}{\rm cm}^{-3}$
is the number density of the ambient medium into which the blast wave expands, and
$\xi_B$ is the energy density of the magnetic field behind the shock in
units of the equipartition value. $\dot{n}$ is normalized in Eq.
(\ref{eq:N_R}) to the value obtained under the assumption that the
GRB rate follows the redshift evolution of star formation rate (\cite{Schmidt01}).
If the GRB rate does not evolve with redshift, $\dot{n}$ is a factor of 8 larger
than the value used in Eq. (\ref{eq:N_R}).
Since afterglow observations generally imply
$\xi_B n\ge 10^{-3}{\rm cm}^{-3}$,
and values $\xi_B n\sim 1{\rm cm}^{-3}$ are obtained in several cases
[see discussion in the paragraphs following
Eq. (\ref{N})], our choice of $n$ and $\xi_B$ values are conservative.
Even for such a conservative choice of parameters we expect $\approx20$
radio afterglows to be detected by an all sky survey
with $f_{\nu \rm min}\simeq5$~mJy. These sources are detected out to a distance
\begin{equation}
d_{\rm max}\approx 200 n_{-1}^{3/8}\left(\frac{\xi_B}{0.03}\right)^{3/8}
\left(\frac{f_{\nu min}}{5\rm mJy}\right)^{-1/2}{\rm Mpc}.
\label{eq:d_max}
\end{equation}

We conducted a search for point-like radio transients with flux
densities larger than 6 mJy using the FIRST
and NVSS surveys, and discovered 25 radio afterglow candidates, listed
in Table 1. The two main
types of sources that may produce radio transients, which may be confused in
our analysis with a radio afterglow, are radio supernovae (RSNe) and radio
loud AGNs. While we have shown that our candidate sources are unlikely to be
RSNe (see sec. \S\ref{sec:RSNe}), the possibility that most of them are radio loud
AGNs cannot be ruled out (see sec. \S\ref{sec:AGN}).
Thus, our detected sources allow to set an upper
limit to the number of orphan radio afterglows. Correcting for various
completeness factors (see sec. \S\ref{sec:Analysis}),
we obtained an upper limit of 447 all sky radio afterglows
above 6 mJy, and 154 above 10 mJy.

We have shown that
for a given isotropic equivalent burst energy, $E_{\rm iso}=f_b^{-1}E$, the
number of afterglows detected in a flux limited survey is smaller for
larger beaming factor $f_b^{-1}$, i.e. for smaller jet opening angle
$\theta$ [see Eq. (\ref{N})].
The upper limit we derived on the number of radio afterglow sources therefore
implies a lower limit for the beaming factor,
\begin{equation}
f_b^{-1}\ge 10\left(\frac{\dot{n}}{0.5{\rm Gpc}^{-3}{\rm yr}^{-1}}\right)^{6/5}
\left(\frac{\xi_B}{0.03}\right)^{27/20}n_{-1}^{19/20}.
\label{f>}
\end{equation}
Here we have used Eq. (\ref{N}) with
$<E_{\rm iso}^{11/6}>^{6/11}=5\times10^{53}$~erg, based on GRBs with known
redshifts (Frail et al. 2001). This result is consistent with the value
$<f_b^{-1}>=500$ inferred by Frail et al. (2001).
Better determination of the distribution of $n$ and $\xi_B$,
based on afterglow observations of identified GRBs, will improve the
constraint on beaming. Alternatively, if the beaming factor
is determined independently using other methods, then eq. (\ref{f>}) can be
used to constrain
the afterglow parameters.  For instance, invoking $<f_b^{-1}>=500$ implies
\begin{equation}
\left(\frac{\xi_B}{0.03}\right)^{27/19}n<5
\left(\frac{\dot{n}}{0.5{\rm Gpc}^{-3}{\rm yr}^{-1}}\right)^{-24/19}{\rm cm}^{-3}.
\label{eq:xiBn}
\end{equation}
In reality
the afterglow parameters in a sample of radio GRBs should have a spread, and so
the letter condition may be taken as a constraint on the mean.

Additional observational efforts aimed at identifying the  candidates in table 1
are strongly motivated by two arguments. First, some of the candidates may turn
out to be orphan afterglows.
The fact that most candidates in table 1 have no optical counterparts brighter
than the POSS-II limit, implies that if some of them are indeed GRB radio orphans
then their host galaxies are not as bright as would have been expected under the
assumption that GRBs are associated with star-forming regions, given the distances
estimated in Eq. (\ref{eq:d_max}) for the conservative values of $n$ and $\xi_B$.
Nevertheless, the observed number of orphan candidates is consistent with
larger values of these parameters, as shown in Eq. (\ref{eq:xiBn}), for which the
distances are larger and the absence of optical counterparts is not surprising.
Second, further observations are also motivated by the fact that rejection of
sources from our list in Table 1 as possible candidates would impose a
more stringent constraint for the beaming factor. Rejection of all the sources in
Table 1 would imply an all sky upper limit of 49 sources, and a corresponding
beaming factor $f_b^{-1}>100$.

Future radio surveys may go deeper than the FIRST and would be of
comparable or better quality.
%Eli 14/3
The Allen Telescope Array (formerly the "1-hectar radio
telescope") will allow a full coverage of the northern sky every
night with $\sim1$~mJy sensitivity\footnote{see
http://astron.berkeley.edu/ral/}, and the square-kilometer-array
may provide even better
sensitivity\footnote{http://www.nfra.nl/skai/}.
%-----
A search for point-like radio transients similar to that outlined
in \S\ref{sec:Analysis}, would probably yield a large number of
candidates. The analysis presented in \S\ref{sec:Expected}
predicts that the number of all sky radio GRBs above 1 mJy is
likely to lie in the range between a few hundreds and a few
thousands, depending on afterglow parameters [see Eq.
(\ref{eq:N_R})], provided the average beaming factor of GRBs is
not significantly larger than currently estimated from other
considerations.
%Eli 14/3
Since the detectable orphan population is dominated by $\sim1$~yr
old GRB remnants, which have just undergone the transition to
sub-relativistic spherical expansion, these sources should be
transient, exhibiting $\sim1/t$ flux decline on time scale of
months. Expansion at the speed of light over $\sim1$~yr implies
that the angular size of the nearby remnants, lying at a distance
$\sim100$~Mpc, is $\approx1$~mas. The nearby remnants may
therefore be resolved using the VLBA and, moreover, since they
have just undergone the transition to spherical expansion, their
structure may show significant anisotropy (see Paczy\'nski 2001).
%-----

RSNe and radio loud AGNs would probably contribute a large number
of transients, and may render the identification of radio
afterglows difficult.  As shown in \S\ref{sec:RSNe}, RSNe and
RGRBs can, in principle, be distinguished statistically by their
redshift distributions. At the mJy level, the average distance of
the radio afterglows is expected to lie between 0.1 to 1 Gpc,
depending on afterglow parameters [see Eq. (\ref{eq:d_max})], so
it should be possible to detect the host galaxies and obtain their
redshifts. RSNe should typically lie at significantly shorter
distances (see \S\ref{sec:RSNe}). Detection of such bimodality in
redshift distribution will greatly assist in identifying radio
GRBs.
%Eli 14/3
Furthermore, the relativistic expansion of the GRB remnants also
implies, as explained above, that unlike RSNe they may be resolved
with the VLBA, possibly revealing anisotropic structure.
%----------------
Efficient rejection of flaring blazars would require follow-up
observations in the radio, optical and X-ray bands. In particular,
the orphan radio afterglows should exhibit a $t^{-1}$ decay of the
radio flux.

Finally, it is conceivable that GRBs emit substantial amounts of
energy in the form of gravitational waves and high-energy
neutrinos (\cite{WnB97,RnM98,WnB99,WnB-AGnus,MnW01}; see
\cite{W01Rev} for review). Association of orphan radio afterglows
with potential signals in gravitational wave and neutrino
detectors may help increasing the signal-to-noise ratio. In a
model proposed recently (\cite{VP01,VPL02}), the GRB results from
extraction of the rotational energy of a Kerr black hole through
its interaction with a magnetized torus.  This model predicts that
the major fraction of the rotational energy available will be
emitted in the form of gravitational waves in the band 0.5 - 1.5
KHz, for a reasonable range of black hole masses.  It further
proposes that measuring the product of the total energy carried by
the gravitational waves and the corresponding frequency, which
reflects the compactness of the system, can provide an existence
test for Kerr black holes. However, such a measurement requires a
knowledge of the distance to the source.  Since the gravitational
wave emission is isotropic, the chance of associating a LIGO
source with a GRB event is small if GRBs have a beaming factor as
large as currently believed.  Nevertheless, if this model is
correct, then LIGO sources should be associated with the orphan
radio afterglows that should appear several months after the burst
of gravitational radiation.  At the distance within which LIGO can
detect the predicted signal ($\sim$ 100 Mpc), the radio source is
expected to be very bright (see eq. [\ref{fmax}]) and can be
monitored for at least several years.  Even though the angular
resolution of the gravitational wave telescopes is expected to be
quite poor, the chance that such a bright transient would coincide
with a LIGO source should still be very small, and so such an
association would be statistically significant.  Our analysis for
instance yielded an upper limit of about $10^{-2}$ transients per
square degree above 10 mJy.  The association of a radio afterglow
with a LIGO source would also provide a means to identify the host
galaxy, given the superior resolution of the radio telescopes, and
hence a distance measurement for the LIGO source.

%Eli 6/3
\acknowledgements
%Eli 14/3
We thank K. Weiler, A. Loeb B. Paczy\'nski and R.D. Blandford for useful
discussions.
%-----
EW is partially supported by AEC grant and a Minerva
grant, and is an incumbent of the Beracha Career development
Chair.
%-----

\break

\end{document}